\documentclass[12pt]{article}
\usepackage{amsmath}
\usepackage{amsthm}
\usepackage{amsfonts}

\newtheorem{theorem}{Theorem}
\newtheorem{definition}{Definition}

\title{Complex structure of a real Clifford algebra}
\author{jason hanson}
\date{}

\begin{document}
\maketitle
\abstract{The classification of real Clifford algebras in terms of matrix algebras is well--known.  Here we consider the real Clifford algebra ${\mathcal Cl}(r,s)$ not as a matrix algebra, but as a Clifford module over itself.  We show that ${\mathcal Cl}(r,s)$ possesses a basis independent complex structure only when the square of the volume element $\omega$ is $-1$, in which case it is uniquely given up to sign by right multiplication with $\omega$.}

\section{Introduction}

\begin{definition}\label{def:complex}
Let ${\mathcal A}$ be an associative algebra over ${\mathbb R}$ acted on by a group $G$.  An {\bf equivariant complex structure} on ${\mathcal A}$ is an ${\mathbb R}$--linear map $J:{\mathcal A}\rightarrow{\mathcal A}$ such that (i) $J^2=-I$, (ii) $J(xy)=xJ(y)$ for all $x,y\in{\mathcal A}$ , and (iii) $J(g\cdot x)=g\cdot J(x)$ for all $g\in G$ and $x\in{\mathcal A}$.
\end{definition}

We use the notational conventions of \cite{Lawson} (which also serves as a blanket reference), namely ${\mathcal Cl}(r,s)$ denotes the Clifford algebra over ${\mathbb R}^{r+s}={\mathbb R}^r\oplus{\mathbb R}^s$ with inner product $q=I\oplus(-I)$ and Clifford condition ${\bf u}^2=-{\bf u}^Tq{\bf u}$, for ${\bf u}\in{\mathbb R}^{r+s}$.  The symbols $e_1,\dots,e_{r+s}$ denote the standard basis of ${\mathbb R}^{r+s}$, and the volume element of ${\mathcal Cl}(r,s)$ is $\omega\doteq e_1\cdots e_{r+s}$.

The Lie group $O(r,s)$ consists of linear transformations on ${\mathbb R}^{r+s}$ that preserve the inner product: $\Lambda^Tq\Lambda=q$.  The $O(r,s)$ action extends to an action on ${\mathcal Cl}(r,s)$ via the requirement $\Lambda\cdot(xy)=(\Lambda\cdot x)(\Lambda\cdot y)$ for all $x,y\in{\mathcal Cl}(r,s)$.  The subgroup $SO(r,s)$ consists of elements of $O(r,s)$ of unit determinant.  We have that $\Lambda\cdot\omega=(\det\Lambda)\omega$ for all $\Lambda\in O(r,s)$.  In particular, $\Lambda\cdot\omega=\omega$ for all $\Lambda\in SO(r,s)$.

In the case when the algebra in definition \ref{def:complex} is ${\mathcal Cl}(r,s)$, condition (ii) implies that a complex structure, if it exists, is completely determined by the value $J(1)$, since $J(x)=J(x1)=xJ(1)$.  Moreover, the complex structure is $SO(r,s)$--equivariant if and only if $J(1)$ is invariant; i.e., $\Lambda\cdot J(1)=J(1)$ for all $\Lambda\in SO(r,s)$.  Indeed, if $J(1)$ is invariant, then $\Lambda\cdot J(x)=\Lambda\cdot(xJ(1))=(\Lambda\cdot x)(\Lambda\cdot J(1))=(\Lambda\cdot x)J(1)=J((\Lambda\cdot x)1)=J(\Lambda\cdot x)$.  Conversely, if $J$ is equivariant, then $\Lambda\cdot J(1)=J(\Lambda\cdot 1)=J(1)$.  Note that $SO(r,s)$--equivariance implies that the complex structure is independent of the choice of basis, as long as we restrict ourselves to orthonormal bases with the same orientation.  Our goal is to show the following.

\begin{theorem}\label{thm:main}
The real Clifford algebra ${\mathcal Cl}(r,s)$ possesses a $SO(r,s)$--equivariant complex structure $J$ if and only if either (1) $s$ is odd and $r+s\equiv 0,3\bmod{4}$, or (2) $s$ is even and $r+s\equiv 1,2\bmod{4}$.  Moreover, the only two equivariant complex structures satisfy $J(1)=\pm\omega$.
\end{theorem}

We have that $\omega^2=\pm 1$, and it is well--known that $\omega^2=-1$ precisely when either condition (1) or (2) is met.  Thus the existence of an equivariant complex structure is equivalent to $\omega^2=-1$.

The complex structure considered here is different from the algebraic structure used in the classification of real Clifford algebras.  In the latter, ${\mathcal Cl}(r,s)$ is identified with a matrix algebra, or the sum of two matrix algebras, over either the reals, complex numbers, or quaternions.  In contrast, we view ${\mathcal Cl}(r,s)$ as a module over itself.  That is, we identify ${\mathcal Cl}(r,s)$ with ${\mathbb R}^N$, where $N=2^{r+s}$.  Clifford multiplication then gives a real linear representation $\rho_0:{\mathcal Cl}(r,s)\times{\mathbb R}^N\rightarrow{\mathbb R}^N$, where $\rho_0(x)y\doteq xy$.  Using the induced action of $SO(r,s)$ on ${\mathcal Cl}(r,s)$, we have another representation: $\rho_1:SO(r,s)\times{\mathbb R}^N\rightarrow{\mathbb R}^N$, given by $\rho_1(\Lambda)y\doteq\Lambda\cdot y$.  The complex structure in theorem \ref{thm:main} is then a real linear map $J:{\mathbb R}^N\rightarrow{\mathbb R}^N$ compatible with both representations: $J\rho_x=\rho_xJ$ and $J\rho_\Lambda=\rho_\Lambda J$, where $\rho_xy\doteq\rho_0(x,y)$ and $\rho_\Lambda y\doteq\rho_1(\Lambda,y)$.

For instance, according to the classification of real Clifford algebras, ${\mathcal Cl}(2,0)$ is isomorphic to the quaternions, while ${\mathcal Cl}(0,2)$ is isomorphic to the algebra of $2\times 2$ matrices over the reals.  Under our scheme, both admit complex structures.  In particular for ${\mathcal Cl}(0,2)$, if we take $J(1)=e_1e_2$, then we may identify ${\mathbb R}^4={\mathbb R}\{1,e_1e_2,e_1,e_2\}$ with ${\mathbb C}^2={\mathbb C}\{1,e_1\}$, where $i=e_1e_2$ and $ie_1=e_2$.  This gives us the complex representation of ${\mathcal Cl}(0,2)$ generated by $\rho_{e_1}=\bigl(\begin{smallmatrix}0 & 1\\ 1 & 0\end{smallmatrix}\bigr)$ and $\rho_{e_2}=\bigl(\begin{smallmatrix} 0 & -i\\i & 0\end{smallmatrix}\bigr)$.

\section{Proof of main theorem}

If we take $J(1)=\pm\omega$, then $J^2(1)=J(\pm\omega)=\pm\omega J(1)=\omega^2$.  Thus a complex structure on ${\mathcal Cl}(r,s)$ exists if $\omega^2=-1$.  Moreover, $\omega$ is $SO(r,s)$--invariant, whence $J$ is equivariant.  So to prove theorem \ref{thm:main}, we must show that (1) if $\omega^2=-1$, then the only equivariant complex stricture up to sign is the one thus defined, and (2) if $\omega^2=1$, then an equivariant complex structure does not exist.

Instead of the action on ${\mathcal Cl}(r,s)$ by the Lie group $SO(r,s)$, it is is more convenient to consider its corresponding Lie algebra action:
$$so(r,s)\times{\mathcal Cl}(r,s)\rightarrow{\mathcal Cl}(r,s),
  \quad
  L\cdot(xy)=(L\cdot x)y+x(L\cdot y)
$$
for all $L\in so(r,s)$ and $x,y\in{\mathcal Cl}(r,s)$.  Recall that $so(r,s)$ consists of all linear transformations on ${\mathbb R}^{r+s}$ such that $L^Tq+qL=0$.  In other words, $qL$ is antisymmetric.

Let $E_{jk}$ denote the matrix whose elements are zero unless except at entry $jk$, which has the value of unity.  That is, $E_{jk}e_l=\delta_{kl}e_j$.  For $j<k$, set $L_{jk}\doteq(E_{jk}-E_{kj})q$.  Then $L_{jk}\in so(r,s)$, and one computes
\begin{equation}\label{eq:L0}
  L_{jk}e_j=-\varepsilon_je_k,
  \quad L_{jk}e_k=\varepsilon_ke_j,
  \quad\text{$L_{jk}e_l=0$ if $l\neq j,k$},
\end{equation}
where $\varepsilon_j\doteq 1$ if $1\leq j\leq r$, and $\varepsilon_j\doteq -1$ if $r<j\leq r+s$ (so that $e_j^2=-\varepsilon_j$ and $qe_j=\varepsilon_je_j$).

Now let $I=(i_1,\dots,i_k)$ be such that $0<i_1<\cdots<i_k\leq r+s$, and set $|I|\doteq k$.  Then the collection of all $e_I\doteq e_{i_1}\cdots e_{i_k}$ with $0\leq|I|\leq r+s$ gives the usual vector space basis of ${\mathcal Cl}(r,s)$, provided we set $e_I=1$ if $|I|=0$.  If $|I|=r+s$, then $e_I=\omega$.  When $j\in\{i_1,\dots,i_k\}$, we will write $j\in I$.  The following facts will be of use.
\begin{enumerate}
  \item\label{eq:L1} If $j,k\in I$ or $j,k\not\in I$, then $L_{jk}\cdot e_I=0$
  \item\label{eq:L2} The map $e_I\mapsto L_{jk}\cdot e_I$ gives, up to sign, a bijection between the set of all basis vectors $e_I$ with $j\in I,k\not\in I$ and the set of all basis vectors $e_J$ with $j\not\in J,k\in J$.
\end{enumerate}
In particular, $L_{jk}\cdot e_I\neq 0$ if exactly one of $j,k$ is in $I$.  Observe that $L_{jk}\cdot e_je_k=(L_{jk}\cdot e_j)e_k+e_j(L_{jk}\cdot e_k)=-\varepsilon_je_k^2+\varepsilon_ke_j^2=\varepsilon_j\varepsilon_k-\varepsilon_k\varepsilon_j=0$.  Facts \ref{eq:L1} and \ref{eq:L2} follow from this and  \eqref{eq:L0}.   

Write $J(1)=\sum_I\kappa_Ie_I$, where the sum is over all $I$ with $0\leq|I|\leq r+s$.  We first show that $\kappa_I=0$ unless $|I|=0$ or $|I|=r+s$.  By the invariance of $J(1)$, we must have $L\cdot J(1)=0$ for all $L\in so(r,s)$.  In particular, we have ($\ast$) $\sum_I\kappa_IL_{jk}\cdot e_I=0$ for all $j<k$.  By facts \ref{eq:L1} and \ref{eq:L2}, the only nonzero summands are those with index $I$ where $j\in I$, $k\not\in I$ or $j\not\in I$, $k\in I$.  Moreover for such indices $I$, the basis vectors $L_{jk}e_I$ are linearly independent, and so ($\ast$) implies $\kappa_I=0$.  However, if $0<|I|<r+s$, we can always find $j<k$ such that $j\in I$, $k\not\in I$ or $j\not\in I$, $k\in I$.  Consequently, ($\ast$) implies that $\kappa_I=0$ if $0<|I|<r+s$.  That is, $J(1)=c+d\omega$ for some real numbers $c,d$.  We show that $c=0$ and $d=\pm 1$.  Indeed, $-1=J(1)^2=(c^2+d^2\omega^2)+2cd\omega$.  Again by linear independence, $c^2+d^2\omega^2=-1$ and $cd=0$.  The only a real solution to these equations occurs when $c=0$, $\omega^2=-1$, and $d=\pm 1$.

\section{Application to gamma matrices}

According to the classification scheme for Clifford algebras, ${\mathcal Cl}(3,1)$ is isomorphic to the algebra of real $4\times 4$ matrices, while ${\mathcal Cl}(1,3)$ is isomorphic to the algebra of $2\times 2$ quaternionic matrices.  However, both admit equivariant complex structures in the sense of theorem \ref{thm:main}.  For ${\mathcal Cl}(3,1)$, if we use the complex structure with $J(1)=e_{1234}$, then we may write
$${\mathcal Cl}(3,1)
  ={\mathbb C}\{1,e_{12},e_{13},e_{14},e_1,e_2,e_3,e_4\}
$$
(note that is not the same as complexification).  Moreover as a complex Clifford module of dimension eight, ${\mathcal Cl}(3,1)$ splits into two equivalent irreducible summands of dimension four.  Indeed, note that if $P:{\mathcal Cl}(3,1)\rightarrow{\mathcal Cl}(3,1)$ is a projection onto an irreducible summand, then $P(xy)=xP(y)$ for all $x,y\in{\mathcal Cl}(3,1)$ and $P^2=P$; i.e., $P$ is determined by $P(1)$, where $P(1)$ satisfies $P(1)^2=P(1)$.  However, $P$ is not unique and we do not require that $P$ be $SO(3,1)$--equivariant (in fact, one may show that the only equivariant projection is the identity).  A possible projection is given by $P(1)=\tfrac{1}{2}(1-e_{14})$, in which case,
$${\rm im}(P)={\mathbb C}\{1-e_{14},e_{12}-ie_{13},e_1+e_4,e_2-ie_3\}.$$
As a ${\mathcal Cl}(3,1)$--module, the effect of Clifford multiplication by the algebra generators $e_1,e_2,e_3,e_4$ with elements of ${\rm im}(P)$ is multiplication by the $4\times 4$ complex matrices
$$\gamma_1=\begin{pmatrix}
             0 & -\sigma_3\\
             \sigma_3 & 0
           \end{pmatrix},
  \gamma_2=\begin{pmatrix}
             0 & -\sigma_1\\
             \sigma_1 & 0
           \end{pmatrix},
  \gamma_3=\begin{pmatrix}
             0 & -\sigma_2\\
             \sigma_2 & 0
           \end{pmatrix},
  \gamma_4=\begin{pmatrix}
             0 & I\\
             I & 0
           \end{pmatrix}
$$
with $\sigma_j$ ($j=1,2,3$) the $2\times 2$ Pauli spin matrices.


\end{document}